\documentclass[sigplan,screen,nonacm]{acmart}

\usepackage{booktabs}
\usepackage{tabularx}
\usepackage{listings}
\usepackage{subcaption}
\include{url}

\AtBeginDocument{%
  \providecommand\BibTeX{{%
    \normalfont B\kern-0.5em{\scshape i\kern-0.25em b}\kern-0.8em\TeX}}}

\begin{document}

\title{LeftoverLocals: Listening to LLM Responses Through Leaked GPU Local Memory}

\author{Tyler Sorensen}
\affiliation{%
  \institution{Trail of Bits\\University of California, Santa Cruz}
  \city{Santa Cruz}
  \state{California}
  \country{USA}
}

\author{Heidy Khlaaf}
\affiliation{%
  \institution{Trail of Bits}
  \city{New York City}
    \state{New York}
  \country{USA}
  }


\begin{abstract}
  This paper describes LeftoverLocals: a vulnerability that allows data recovery from GPU memory created by another process on Apple, Qualcomm, and AMD GPUs. LeftoverLocals impacts the security posture of GPU applications, with particular significance to LLMs and ML models that run on impacted GPUs. By recovering local memory – an optimized GPU memory region – we built a PoC where an attacker can listen into another user’s interactive LLM session (e.g., llama.cpp) across process or container boundaries.
\end{abstract}



\setcopyright{none}

\maketitle

\section{Introduction}

%This paper presents a GPU memory leak, dubbed LeftoverLocals, that was disclosed earlier this year (CVE-2023-4969\footnote{\url{https://kb.cert.org/vuls/id/446598}}). This leak allows a co-resident process to read leftover data in an optimized GPU memory region called local memory and impacts GPUs from AMD, Apple, Qualcomm and Imagination (leaking up to 5MB of data on large AMD GPUs). While the leak impacts many GPU applications, we highlight the impact on ML applications, showing that the leaked data is sufficient for an attacker to reconstruct responses from an interactive LLM session being run by a co-resident victim process. Using LeftoverLocals as a case-study, we will discuss various security concerns that can arise when utilizing GPU for general purpose computation, which is the computational workhorse of many ML applications. 

\textit{This paper is essentially a direct port (by the authors) from the official Trail of bits blog post, which can be found here: \url{https://blog.trailofbits.com/2024/01/16/leftoverlocals-listening-to-llm-responses-through-leaked-gpu-local-memory/}. This document was created because an archived paper may be better suited for distribution and citation in some cases.}

Given the rise of ML applications, especially for privacy sensitive application domains, it is imperative to rigorously examine security properties throughout the ML stack. This includes the computational work horse of ML applications, the GPU. While NVIDIA GPUs currently appear to dominate the ML market, most other hardware vendors create their own GPUs, which are becoming more widely used for ML applications, especially for local computation. This work discusses a data leak vulnerability found on many of these GPUs, specifically, from AMD, Apple, Qualcomm and Imagination. This leak, dubbed LeftoverLocals, was disclosed earlier this year (CVE-2023-4969\footnote{\url{https://kb.cert.org/vuls/id/446598}}) and allows a co-resident process to read leftover data in an optimized GPU memory region called local memory.

\begin{figure}
    \centering
    \fbox{\includegraphics[width=0.45\textwidth]{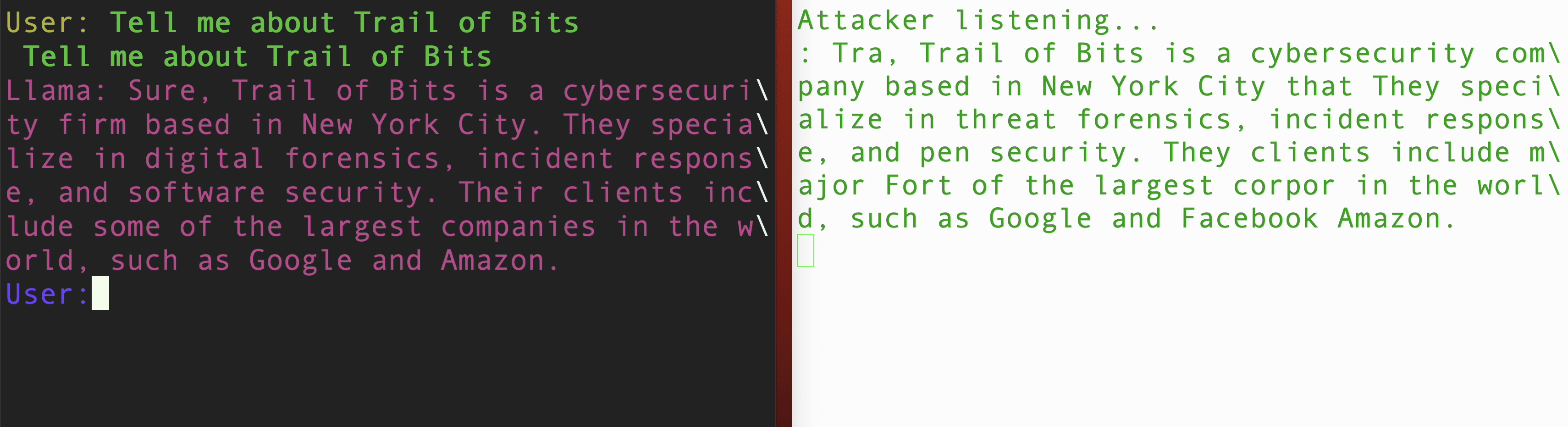}}
    \caption{An example of an LLM response that an attacker was able to reconstruct utilizing LeftoverLocals. The victim terminal is on the left (black background), and the attacker terminal is on the right (white backgroun). We can see that the attacker is able to reconstruct the response with relatively high fidelity. We believe that the listener could be more finally tuned to be even more accurate. Details proof-of-concept attack application, along with the system, can be found in Sec.~\ref{sec:listening}.}
    \label{fig:LLM-leak}
\end{figure}

LeftoverLocals can leak approximately ~5.5 MB per GPU invocation on an AMD Radeon RX 7900 XT which, when running a 7B model on llama.cpp, adds up to ~181 MB for each LLM query. This is enough information to reconstruct the LLM response with high precision, an example is shown in Fig.~\ref{fig:LLM-leak}. The vulnerability highlights that many parts of the ML development stack have unknown security risks and have not been rigorously reviewed by security experts.

Trail of Bits worked with CERT Coordination Center on a large coordinated disclosure effort involving all major GPU vendors, including: NVIDIA, Apple, AMD, Arm, Intel, Qualcomm, and Imagination. As of writing (Feb. 1, 2024), the status of the impacted vendors, Apple, AMD, and Qualcomm are as follows:

\begin{itemize}
\item \textbf{Apple}: Despite multiple efforts to establish contact through CERT/CC, we only received a response from Apple on January 13, 2024. We re-tested the vulnerability on January 10 where it appears that some devices have been patched, i.e., Apple iPad Air 3rd G (A12). However, the issue still appears to be present on the Apple MacBook Air (M2). Furthermore, the recently released Apple iPhone 15 does not appear to be impacted as previous versions have been. Apple has confirmed that the A17 and M3 series processors contain fixes, but we have not been notified of the specific patches deployed across their devices.
\item \textbf{AMD}: We have confirmed with AMD that their devices remain impacted, although they continue to investigate potential mitigation plans. Their statement on the issue can be read here\footnote{\url{https://www.amd.com/en/resources/product-security/bulletin/amd-sb-6010.html}}.
\item \textbf{Qualcomm}: We received notice that there is a patch to Qualcomm firmware v2.07\footnote{\url{https://lore.kernel.org/linux-firmware/20240111114032.126035-1-quic_akhilpo@quicinc.com/}} that addresses LeftoverLocals for some devices. However, there may still be other devices impacted at this time. A Qualcomm representative has provided the following comment: “Developing technologies that endeavor to support robust security and privacy is a priority for Qualcomm Technologies. We commend Dr. Tyler Sorensen and Dr. Heidy Khlaaf from the AI/ML Assurance group at Trail of Bits for using coordinated disclosure practices and are in the process of providing security updates to our customers. We encourage end users to apply security updates as they become available from their device makers.”
\item \textbf{Imagination}: Despite not observing LeftoverLocals ourselves across the Imagination GPUs that we tested (see Sec.~\ref{sec:testing}), Google has confirmed that some Imagination GPUs are indeed impacted. Imagination released a fix in their latest DDK release, 23.3, made available to customers in December 2023\footnote{\url{https://www.imaginationtech.com/gpu-driver-vulnerabilities/}}.
\end{itemize}

A list of tested and impacted devices can be found in Sec.~\ref{sec:testing} Other vendors have provided us the following details:

\begin{itemize}
\item \textbf{NVIDIA}: confirmed that their devices are not currently impacted. One reason for this could be that researchers have explored various memory leaks on NVIDIA GPUs previously, e.g., in~\cite{cuda-leaks}, and thus, they are aware of these types of issues.
\item \textbf{ARM}: also confirmed that their devices are not currently impacted.

\end{itemize}

We did not hear a response from the remaining vendor, \textbf{Intel}, we however, we tested two GPUs from them and did not observe that they were impacted (see Sec.~\ref{sec:testing}).

\section{Exploit Brief}

GPUs were initially developed to accelerate graphics computations. In this domain, performance is critical, and previously uncovered security issues have generally not had any significant consequences on applications. Historically, this entailed that GPU hardware and software stacks iterated rapidly, with frequent major architecture and programming model changes. This has led to complex system stacks and vague specifications. For example, while CPU ISAs have volumes of documentation, NVIDIA simply provides a few short tables\footnote{\url{https://docs.nvidia.com/cuda/cuda-binary-utilities/index.html\#instruction-set-reference}}. This type of vague specification has led to alarming issues, both previously~\cite{gpu-programming-assumptions} and currently, as LeftoverLocals exemplifies.

\subsection{Exploitation requirements}

This is a co-resident exploit, meaning that a threat actor’s avenue of attack could be implemented as another application, app, or user on a shared machine. The attacker only requires the ability to run GPU compute applications, e.g., through OpenCL, Vulkan, or Metal. These frameworks are well-supported and typically do not require escalated privileges. Using these, the attacker can read data that the victim has left in the GPU local memory simply by writing a GPU kernel that dumps uninitialized local memory. These attack programs, as our code demonstrates, can be less than 10 lines of code. Implementing these attacks is thus not difficult and is accessible to amateur programmers (at least in obtaining stolen data). We note that it appears that browser GPU frameworks (e.g., WebGPU) are not currently impacted, as they insert dynamic memory checks into GPU kernels.

Unless the user inspects the application’s low-level GPU source-code, it is not possible for them to uncover if their application is utilizing GPU local memory; this matter is further complicated as the GPU code is often hidden deep in library calls, at low levels of deep software stacks (e.g., for ML). Overall, there are very limited ways to observe that an attacker is currently stealing data, or has stolen data. This attack hinges on the attacker reading uninitialized memory on the GPU, and while this is technically undefined behavior, it is not currently checked dynamically, or logged. Any additional defenses would be quite invasive, e.g., performing code analysis on GPU kernels to check for undefined behavior. We have released a PoC that vulnerability shows how this vulnerability can be exploited\footnote{\url{https://github.com/trailofbits/LeftoverLocalsRelease}}, and Sec.~\ref{sec:listening} describes how it works.

\subsection{Mitigations}

Given the lack of comprehensive patches across impacted GPU vendors, LeftoverLocals can be defended by modifying the source code of all GPU kernels that use local memory. Before the kernel ends, the GPU threads should clear memory (e.g., store 0s) to any local memory memory locations that were used in the kernel. Additionally, the users should ensure the compiler doesn’t remove these memory-clearing instructions away (e.g., by annotating their local memory as volatile), as the compiler may detect that the cleared memory is not used later in the kernel. This is difficult to verify because GPU binaries are typically not stored explicitly, and there are very few GPU binary analysis tools. Because of reasons like this, we note that this mitigation may be difficult for many users, and we discuss this further in Sec.~\ref{sec:testing}.

\section{Background: How GPUs work} 

\begin{figure}
    \centering
    \fbox{\includegraphics[width=0.45\textwidth]{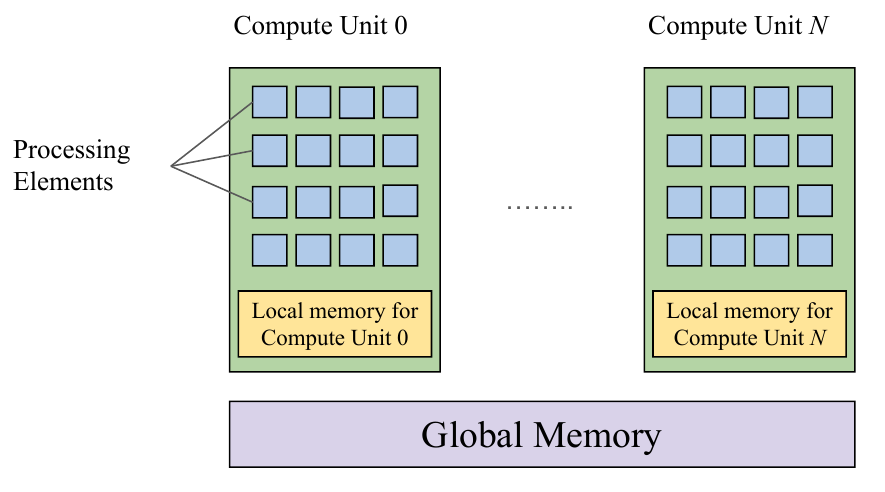}}
    \caption{A simplified view of the GPU architecture: processing elements are partitioned into compute unites. All processing elements have access to global memory (often located in VRAM for discrete GPUs), while only processing elements in the same compute unit share the same local memory.}
    \label{fig:gpu}
\end{figure}

GPUs are massively parallel, throughput-oriented co-processors. While originally designed to accelerate graphics workloads, their design, which balances flexible programming and high computational throughput, has been highly effective in a variety of applications. Perhaps the most impactful current application domain is machine learning, where GPUs are the computational workhorse and are used to achieve nearly all major results in this area.

GPUs are not only in large servers; they are in our phones, our tablets, and our laptops. These GPUs come from a variety of vendors, with almost all major hardware vendors (Apple, AMD, Arm, Qualcomm, Intel, and Imagination) producing their own GPU architecture. These GPUs are increasingly used for ML tasks, especially because doing ML locally can preserve users’ privacy, achieve lower latency, and reduce computational burdens on service providers.

\paragraph{GPU architecture} GPU architecture has a parallel, hierarchical structure, shown in Fig.~\ref{fig:gpu}. At the top level, a GPU is made up of Compute Units (sometimes called Streaming Multiprocessors in NVIDIA literature). Large, discrete GPUs contain many compute units, and smaller, mobile GPUs have fewer. For example, the large AMD Radeon RX 7900 XT discrete GPU has 84 compute units, while the mobile Qualcomm Adreno 740 GPU has 8. All compute units have access to global memory. On discrete GPUs, global memory is implemented using VRAM; on integrated GPUs, global memory simply uses the CPU’s main memory.

Compute units encapsulate both compute and memory components. Compute units contain an array of processing elements; these simple cores are the fundamental units of computation and execute a stream of GPU instructions. In terms of memory, compute units often contain a cache for global memory, but they also contain a special region of memory called local memory. This is an optimized memory region that is shared only across processing elements in the same compute unit. This memory can be accessed with significantly less latency than global memory, but also has much smaller capacity. Different GPUs have varying amounts of local memory per compute unit, typically ranging from 16KB to 64KB. For example, the AMD Radeon RX 7900 XT GPU has 84 compute units and a local memory size of 64KB; thus, the total amount of local memory on the GPU is ~5MB. Local memory is a software-managed cache: the program executing on the processing elements is responsible for loading values into local memory (e.g., values that will be repeatedly used from global memory).

\paragraph{GPU execution model} A GPU program, called a (GPU) kernel, is written in a shader language. Common examples are SPIR-V (Vulkan), OpenCL C, (OpenCL), and Metal Shading Language (Metal). These kernels specify a single entry point function, called the kernel function, which is executed by many invocations (i.e., GPU threads). Invocations have unique built-in identifiers (such as a global ID), which can be used to index a unique data element in a data-parallel program. Invocations are further partitioned into workgroups. Each workgroup is mapped to a compute unit (although many workgroups may execute on the same compute unit, depending on resource requirements). All invocations have access to the same global memory, but only invocations in the same workgroup will share the same local memory.

Applications that use the GPU often launch many short-running kernels. These kernels often correspond to basic operations, such as matrix multiplication or convolution. Kernels can then be executed in sequence; for example, each layer in a deep neural network will be a kernel execution. Local memory is statically allocated at each kernel launch and is not specified to persist across kernel calls.

Platforms generally do not time-multiplex different GPU kernels. That is, if multiple kernels are launched simultaneously (e.g., by different users), the GPU will execute one kernel to competition before the next kernel starts. Because GPU kernels are typically short running, sharing GPU resources at kernel boundaries saves expensive preemption overhead while also maintaining acceptable latency in practice.

\paragraph{Terminology} Because this blog post focuses on portable GPU computing, it uses OpenCL GPU terminology. For readers more familiar with GPU terminology from a different framework (e.g., CUDA or Metal), we provide a translation table in Tab.~\ref{tab:translate}.

\newcolumntype{q}{>{\hsize=.75\hsize}X}

\begin{table}
\small
\caption{A translation table between terminology used in different GPU programming frameworks. This work uses OpenCL terminology, but readers may by more familiar with CUDA or Metal terminology.}
\begin{tabularx}{\columnwidth}{qqX}
\toprule
     \textbf{CUDA} & \textbf{OpenCL} & \textbf{Metal} \\
     \midrule
     thread & work-item & thread \\ 
     thread block & workgroup & threadgroup \\
     shared memory & local memory & threadgroup memory\\
     \bottomrule
\end{tabularx}
\label{tab:translate}
\end{table}

\section{The Vulnerability: LeftoverLocals}

In this section we describe the vulnerability, named LeftoverLocals, in more detail. Section~\ref{sec:testing} details our testing campaign across a wide variety of GPU devices, which found that GPUs from AMD, Apple, and Qualcomm are vulnerable to LeftoverLocals. We also note that while GPU memory leaks are not new, e.g., see~\cite{cuda-leaks}, LeftoverLocals has demonstrated both deeper impact and wider breadth than previously discovered vulnerabilities.

At a high level, we found that several GPU frameworks do not sufficiently isolate memory in the same way that it is traditionally expected in CPU-based frameworks. We have observed that on impacted GPUs, it is possible for one kernel—potentially from another user that is co-resident on the same machine—to observe values in local memory that were written by another kernel. Thus, an attacker who has access to a shared GPU through its programmable interface (e.g., OpenCL) can steal memory from other users and processes, violating traditional process isolation properties. This data leaking can have severe security consequences, especially given the rise of ML systems, where local memory is used to store model inputs, outputs, and weights.

Previous academic work showed that NVIDIA GPUs leaked memory across processes through a variety of memory regions, including local memory and across virtualized environments~\cite{cuda-leaks, virtualized-leak}. However, they examined only GPUs from NVIDIA (and we speculate that the results from this paper may be part of the reason why we didn’t observe LocalLeftovers on NVIDIA GPUs). They also did not discuss the impact on widely deployed use-cases, such as ML. Other works have shown how GPUs leak graphics data, and that a co-resident attacker can reconstruct partial visual information from another process~\cite{stealing-webpages, reconstructing, gpu-zip}. A good survey of existing GPU vulnerabilities can be found in~\cite{security-survey}. Despite these prior works, LeftoverLocals shows that many GPUs remain vulnerable to local memory leaks and that this vulnerability can be exploited in co-resident attacks on important ML applications.

Overall, this vulnerability can be illustrated using two simple programs: a listener and a writer, where the writer stores canary values in local memory, while a listener reads uninitialized local memory to check for the canary values. The listener repeatedly launches a GPU kernel that reads from uninitialized local memory. The writer repeatedly launches a GPU kernel that writes canary values to local memory. Below, we demonstrate how each of these operations is carried out.

\paragraph{The listener}

The listener launches a GPU kernel that reads from uninitialized local memory and stores the result in a persistent main memory region (i.e., global memory). This can be accomplished with the OpenCL kernel shown in Fig.~\ref{fig:listener}.

\begin{figure}
\lstset{numbers=left}
\begin{lstlisting}[language=C, basicstyle=\small,xleftmargin=2em]
__kernel void listener(__global volatile 
                       int *dump) 
{
                       
  local volatile int lm[L_SIZE];
  
  for (int i =  get_local_id(0); 
           i <  L_SIZE; 
           i += get_local_size(0)) 
  {
           
    dump[L_SIZE*get_group_id(0)+i]=lm[i];
  }
}
\end{lstlisting}
\caption{An OpenCL kernel showing how to implement a LeftoverLocals listener. Essentially the kernel dumps uninitialized local memory (from the \texttt{lm} array) into a persistent (global) memory region (in \texttt{dump}) so that it can be examined later by the host. The OpenCL builtin ids (e.g., \texttt{get\_local\_id(0)}), allow the entire local memory dump to be done efficiently in parallel.}
\label{fig:listener}
\end{figure}

The keyword \texttt{\_\_kernel} denotes that this is the GPU kernel function. We pass a global memory array \texttt{dump} to the function. Whatever the kernel writes to this array can be read later by the CPU. We statically declare a local memory array \texttt{lm} with a predefined size \texttt{L\_SIZE} (which we set to be the max size of local memory for each GPU we test). This program technically contains undefined behavior, as it reads from uninitialized local memory. Because of this, we use the \texttt{volatile} qualifier to suppress aggressive compiler optimizations that might optimize away the memory accesses. In fact, the code in the github we provide contains a few more code patterns included to further stop the compiler from optimizing away our memory dump. This process is guided by trial-and-error, rather than a systematic approach.

For each loop iteration, the work-item (thread) reads from a location in local memory, and then stores that value to a unique location in the global \texttt{dump} array. The only tricky part of this code is the indexing, because local memory is disjointed across workgroups, so workgroup local IDs need to be mapped to a unique global ID in dump. The process utilizes OpenCL built-in identifiers to achieve this, which are documented in the OpenCL C specification\footnote{\url{https://registry.khronos.org/OpenCL/specs/3.0-unified/html/OpenCL\_C.html\#work-item-functions}}. At the end of the kernel, \texttt{dump} contains every value that was stored in local memory when the listener kernel started executing. Because \texttt{dump} is in the global memory region, it can be copied back to CPU memory and examined to check for canary values.

\paragraph{The writer}
On the other hand, the writer launches a kernel that writes a canary value to local memory (for example, this work uses the value 123). We show an example of the OpenCL kernel code in Fig.~\ref{fig:writer}.

\begin{figure}
\lstset{numbers=left}
\begin{lstlisting}[language=C, basicstyle=\small,xleftmargin=2em]
__kernel void writer(__global volatile 
                     int *canary) 
{
  local volatile int lm[L_SIZE];
  for (uint i = get_local_id(0); 
            i < L_SIZE; 
            i += get_local_size(0)) 
  {
    lm[i] = canary[i];
  }
}
\end{lstlisting}
\caption{An OpenCL kernel showing how to implement a LeftoverLocals writer. Essentially the kernel writes a canary value to all of local memory (in the \texttt{lm} array) so that a listener can later check to see if it observes canary values. Similar to the listener, the OpenCL builtin ids (e.g., \texttt{get\_local\_id(0)}), allow the writer to write to the entirety of local memory efficiently in parallel.}
\label{fig:writer}
\end{figure}

This code is very similar to the listener, except that rather than dumping local memory, it writes to local memory. In this case,  the code is writing a value from a memory location called \texttt{canary}. We use an extra location so that the compiler does not optimize away the memory write (as it is prone to do with constant values). At the end of the kernel, the writer has filled all available local memory with the canary value.

\paragraph{The Host} The CPU host programs for both the listener and the writer launch their respective kernels repeatedly. In the case of the listener, at each iteration, the CPU analyzes the values observed in the local memory and checks for the canary value. On a server, these two programs can be run by different users or in different Docker containers. On a mobile device, these routines can be run in different apps. The apps can be swapped in and out of focus to alternate reading and writing. \textit{If the listener can reliably read the canary values, then we say that the platform is vulnerable to LeftoverLocals.}

\begin{figure*}[t!]
    \centering
    \begin{subfigure}[b]{0.47\textwidth}
        \centering
        \fbox{\includegraphics[width=.95\textwidth, trim=.5cm 2.5cm 1cm 0.5cm]{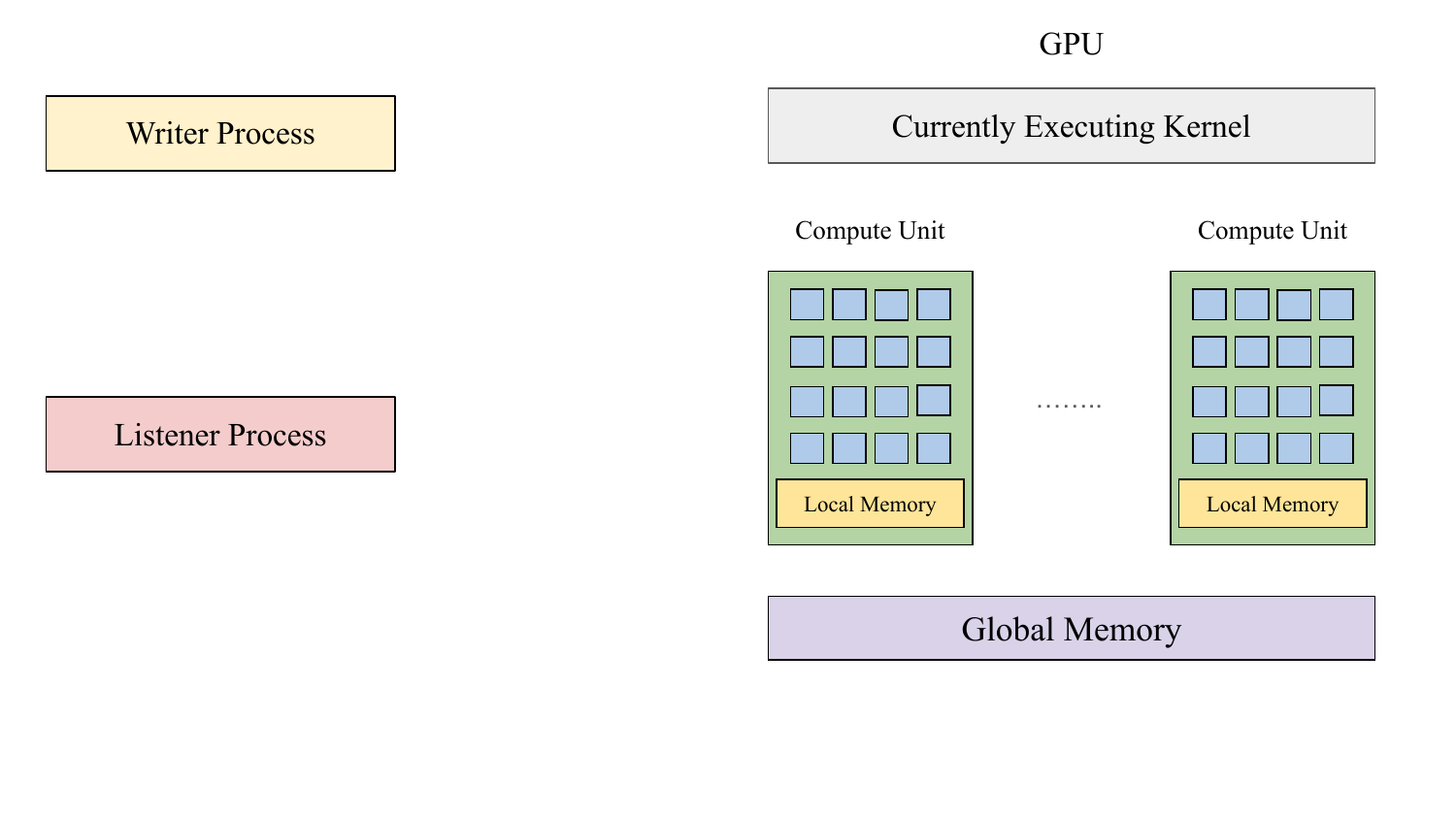}}
        \caption{Initial state of the GPU, where the writer process and the listener process are co-resident, sharing one GPU device}
    \end{subfigure}%
    \hfill
    \begin{subfigure}[b]{0.47\textwidth}
        \centering
       \fbox{\includegraphics[width=.95\textwidth, trim=.5cm 2.5cm 1cm 0.5cm]{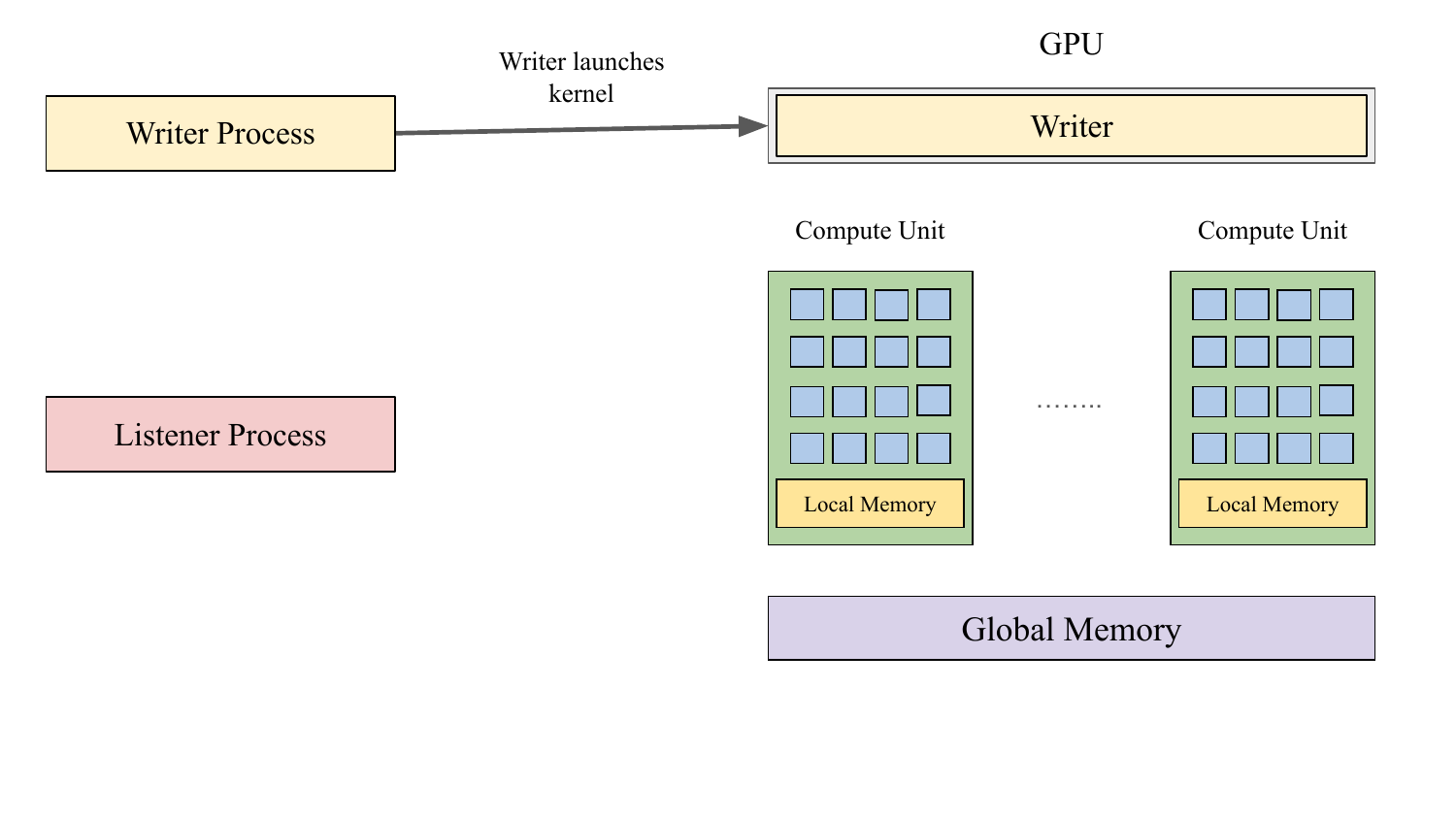}}
        \caption{The writer process launches it's writer kernel, which becomes the sole kernel that is executing on the GPU}
    \end{subfigure}

    \vspace{.4cm}

        \begin{subfigure}[b]{0.47\textwidth}
        \centering
        \fbox{\includegraphics[width=.95\textwidth, trim=.5cm 2.5cm 1cm 0.5cm]{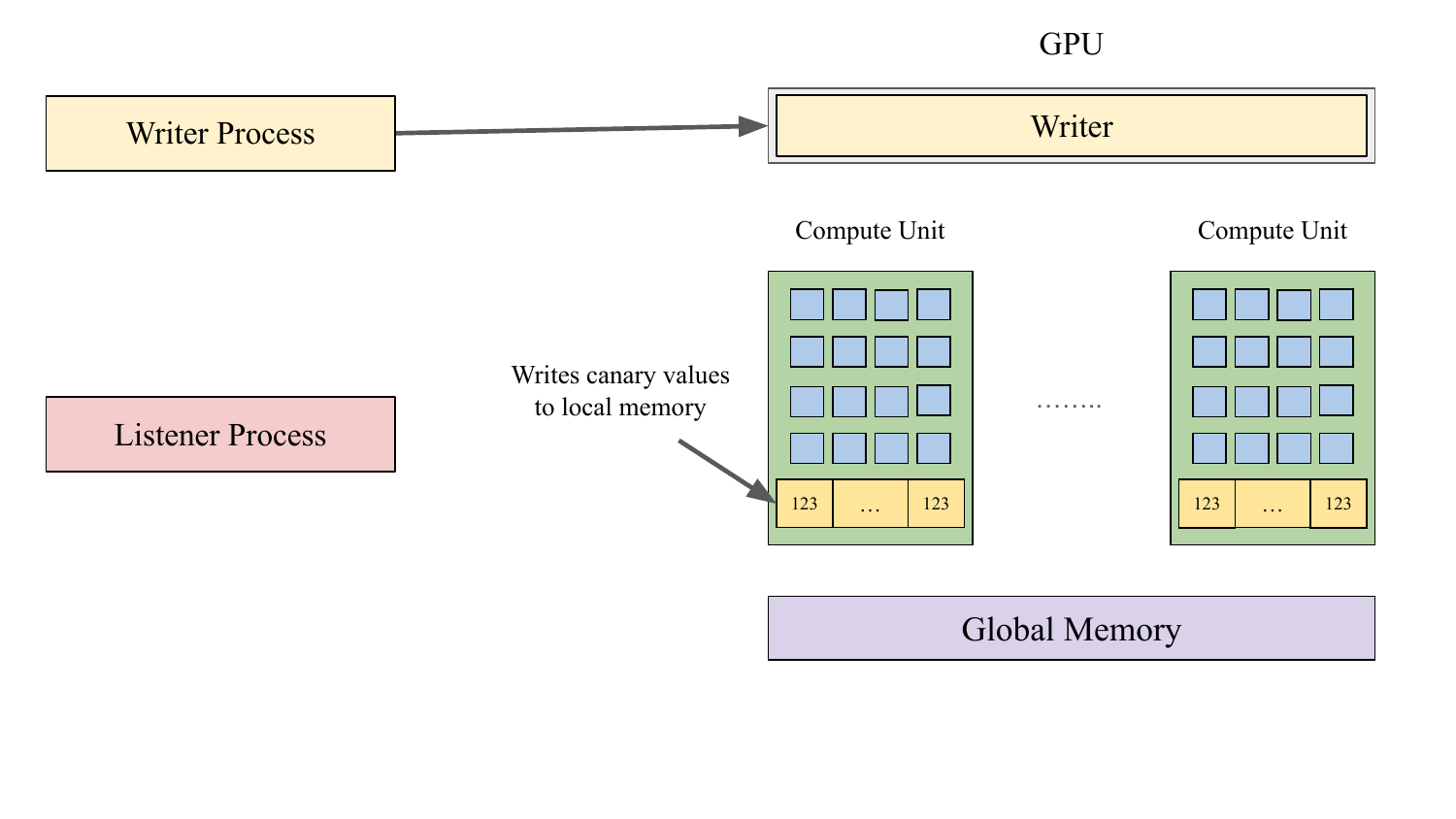}}
        \caption{The writer kernel writes a canary value, in this case 123 to all the local memory locations across all the compute units}
    \end{subfigure}%
    \hfill
    \begin{subfigure}[b]{0.47\textwidth}
        \centering
       \fbox{\includegraphics[width=.95\textwidth, trim=.5cm 2.5cm 1cm 0.5cm]{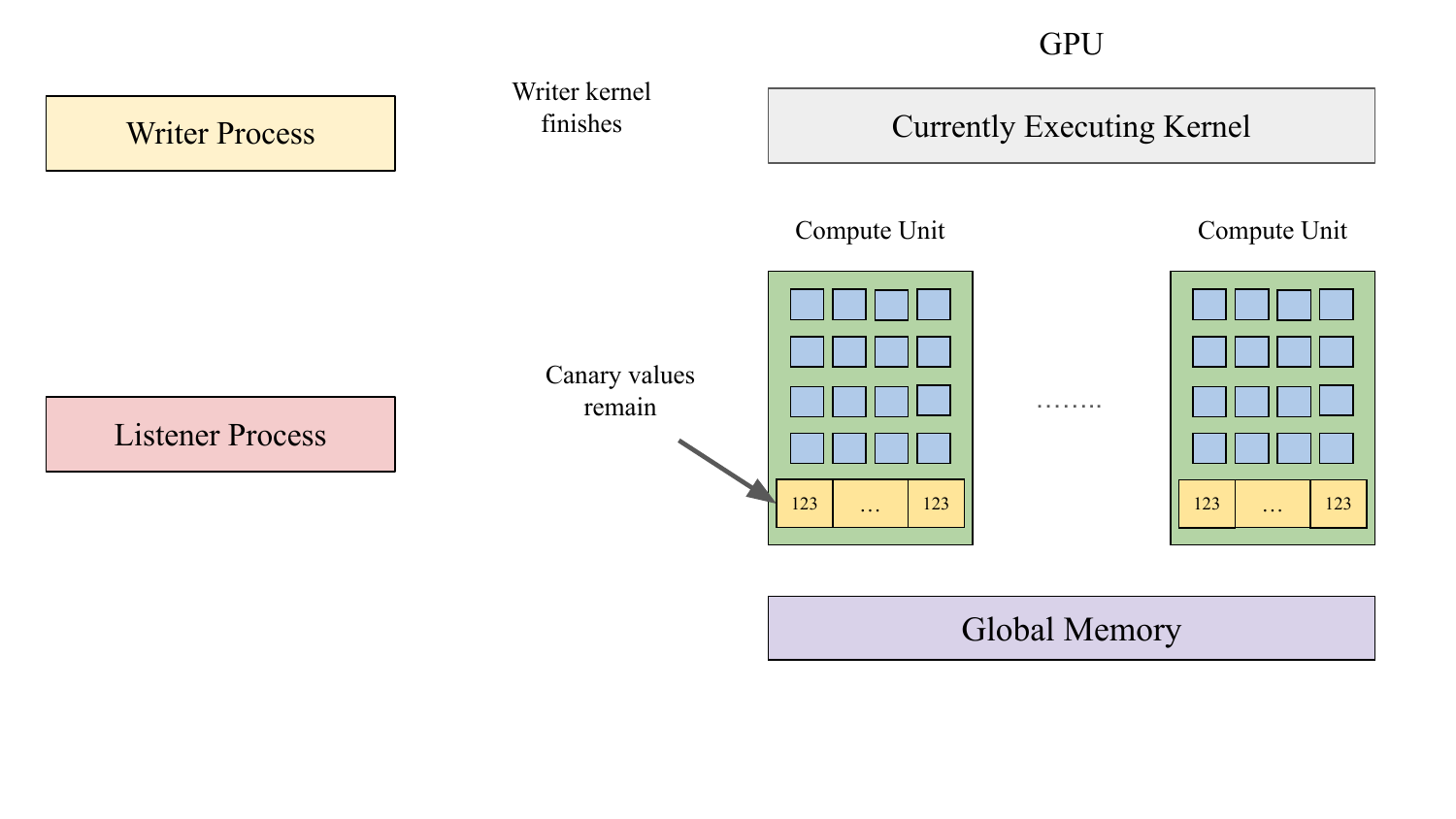}}
        \caption{The writer kernel finishes execution. In a system that is impacted by LeftoverLocals, the canary values stay in local memory}
    \end{subfigure}

        \vspace{.4cm}

       \begin{subfigure}[b]{0.47\textwidth}
        \centering
        \fbox{\includegraphics[width=.95\textwidth, trim=.5cm 2.5cm 1cm 0.5cm]{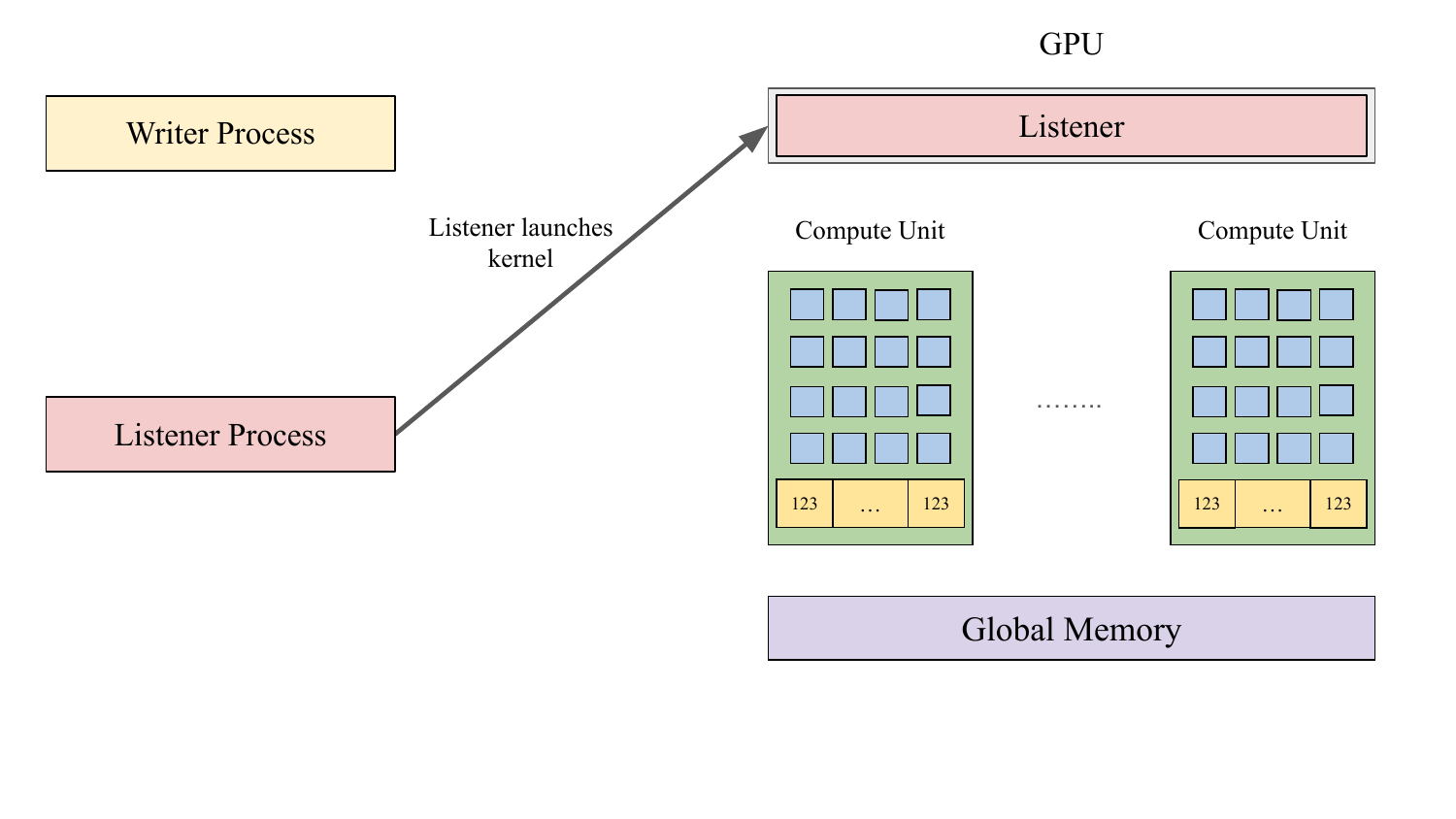}}
        \caption{The listener now gets a turn to execute its kernel, sololy occupying the GPU}
    \end{subfigure}%
    \hfill
    \begin{subfigure}[b]{0.47\textwidth}
        \centering
       \fbox{\includegraphics[width=.95\textwidth, trim=.5cm 2.5cm 1cm 0.5cm]{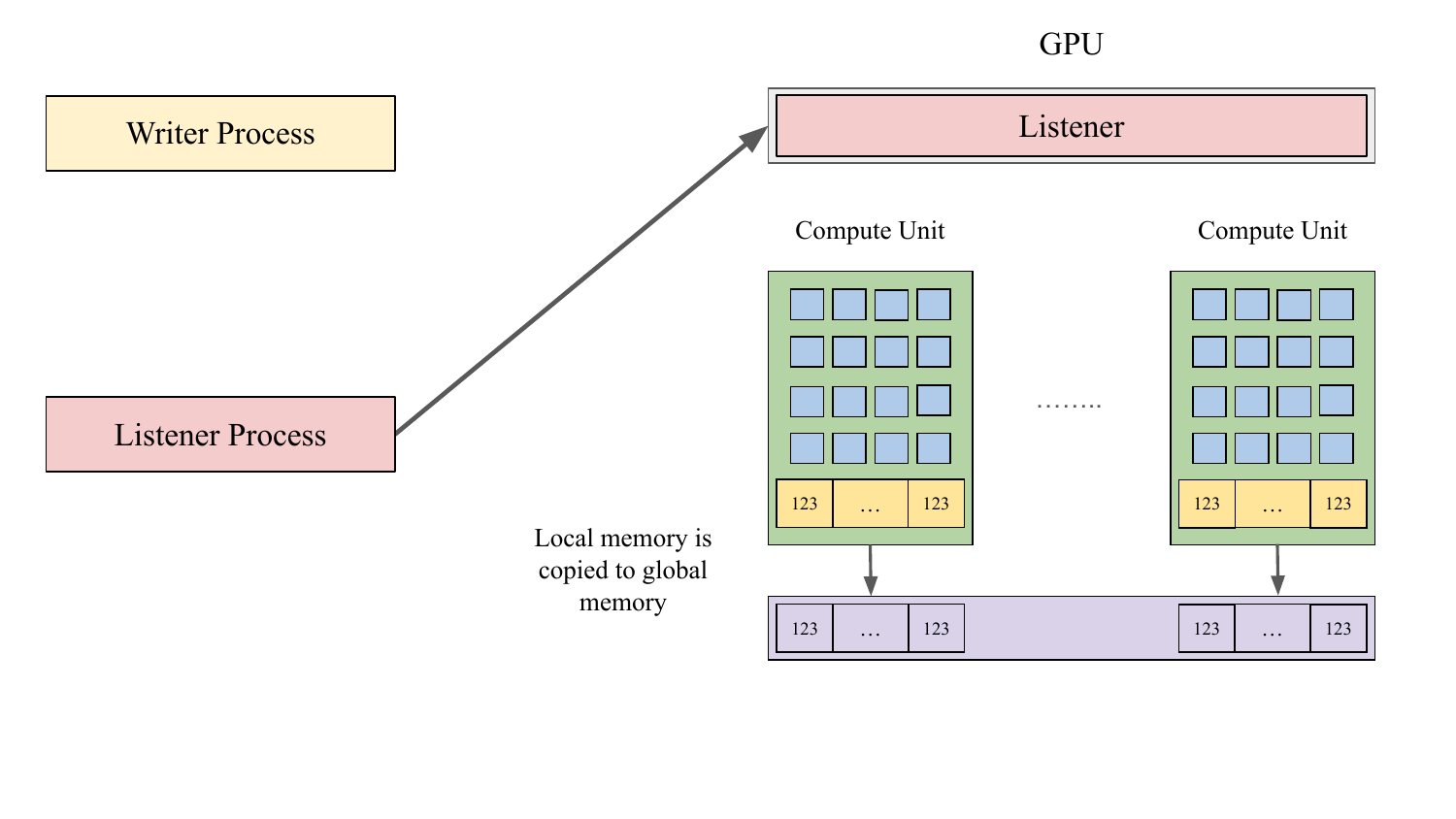}}
        \caption{The listener kernel copies all local memory on the GPU into global memory}
    \end{subfigure}

        \vspace{.4cm}

           \begin{subfigure}[b]{0.47\textwidth}
        \centering
        \fbox{\includegraphics[width=.95\textwidth, trim=.5cm 2.5cm 1cm 0.5cm]{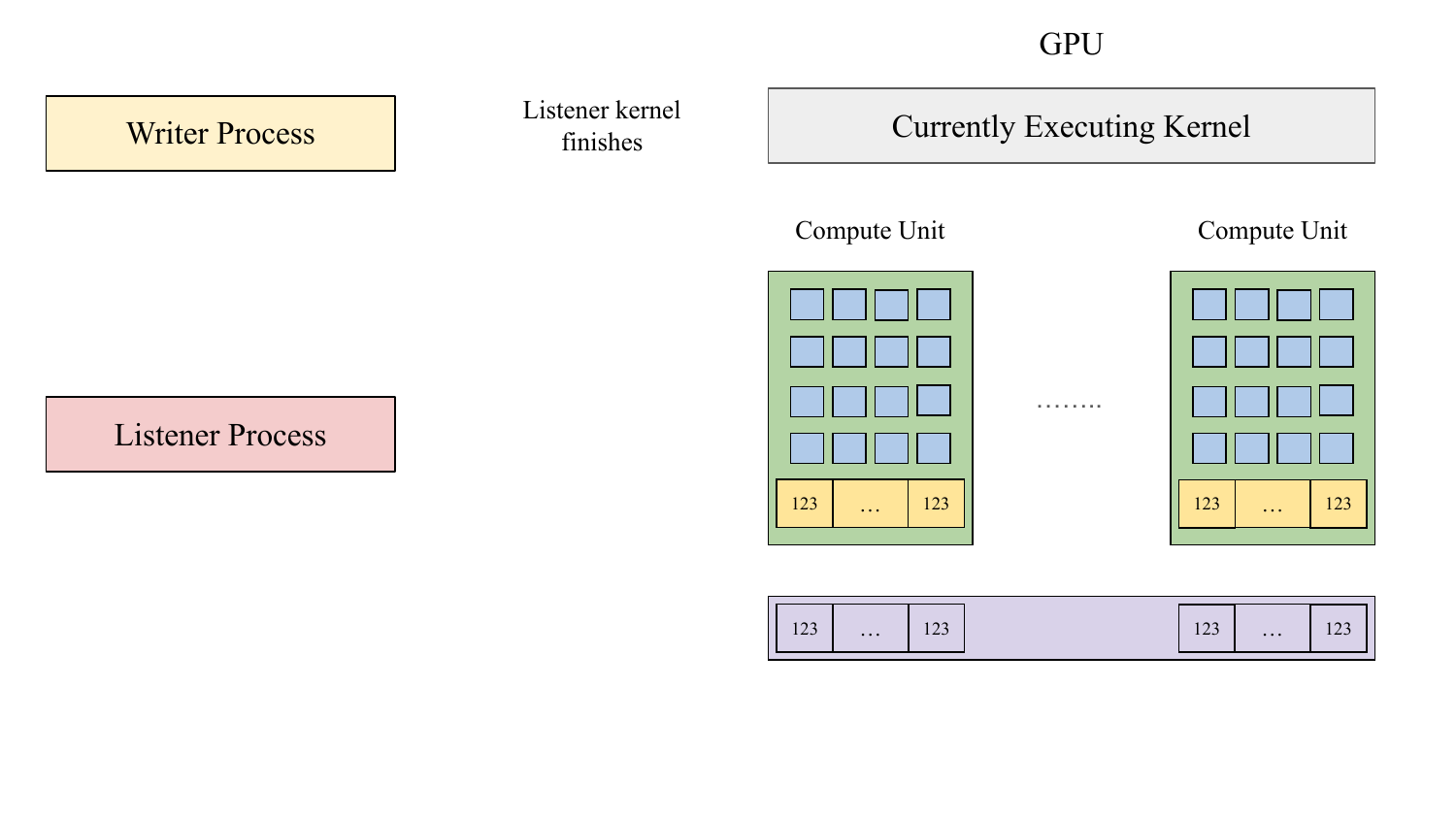}}
        \caption{The listener kernel finishes. At this point, global memory (owned by the listener process) contains a dump of local memory}
    \end{subfigure}%
    \hfill
    \begin{subfigure}[b]{0.47\textwidth}
        \centering
       \fbox{\includegraphics[width=.95\textwidth, trim=.5cm 2.5cm 1cm 0.5cm]{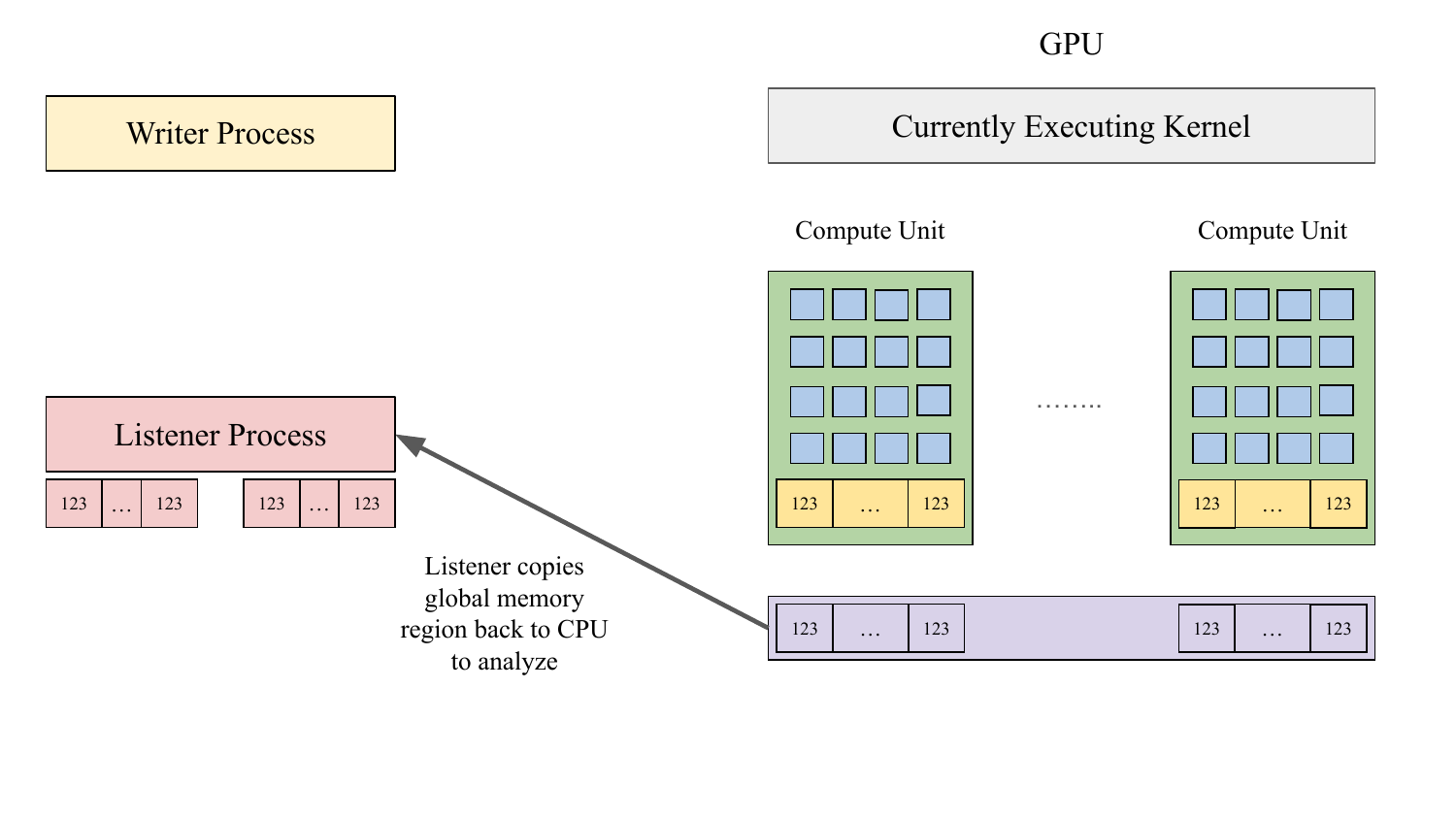}}
        \caption{The listener can now copy global memory back to the CPU and check for leaked data, e.g., canary values}
    \end{subfigure}
    \caption{A series of images illustrating how the listener and the writer interact, and how they can test for the LocalLeftover vulnerability.  \label{fig:series}}
    
\end{figure*}

Figure~\ref{fig:series} shows a series of images on how the listener and the writer interact, and how the listener may observe values from the writer if local memory is not cleared.

\section{Listening to LLM Responses \label{sec:listening}}

\begin{figure*}
    \centering
    \fbox{\includegraphics[width=.95\textwidth]{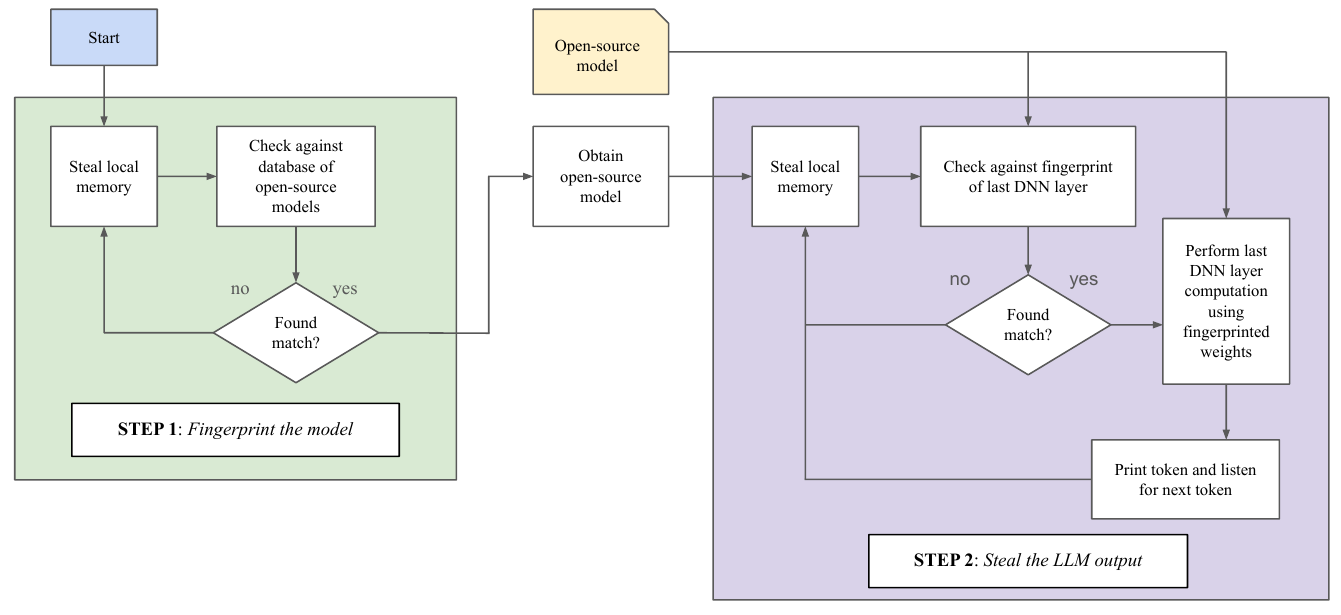}}
    \caption{A flow chart outlining the attack where a malicious listener (the attacker) steals the output of another process (the victim) executing an interactive LLM application. These steps are exclusively followed by the attacker.}
    \label{fig-flowchart}
\end{figure*}

In this section, we provide an overview of how LeftoverLocals can be exploited by a malicious actor (an attacker) to listen to another user’s (the victim) LLM responses on a multi-tenant GPU machine, followed by a detailed description of our proof-of-concept (PoC) attack.

At a high level, both actors are executed as co-resident processes. The attack process implements the listener described above, with the additional steps of comparing the stolen values to various fingerprints. The victim process is unknowingly the writer, where instead of canary values, the values being written are sensitive components of an interactive LLM chat session. The attack ultimately follows two steps, outlined in Fig.\ \ref{fig-flowchart}:

\begin{enumerate}
\item The attack process fingerprints the model that the victim process is using by repeatedly dumping (i.e., stealing) the leftover local memory, which, in this scenario, consists of sensitive components of linear algebra operations used by the victim in the LLM model architecture. This requires that the victim is using an open-source model that the attacker can obtain.
\item The attacker then repeatedly dumps (i.e., steals) the local memory again, specifically searching for the last, i.e., output, layer in the LLM's DNN, which can be identified using weights or memory layout patterns from the earlier fingerprinting.

\end{enumerate}

Note that the output layer is a matrix-vector multiplication with two inputs: the model weights, and the layer input—in other words, the values derived from the user input that propagated through the earlier levels of the deep neural network (DNN). Given that the model weights of the output layer are too large to comprehensively steal, an attacker can inspect available open-source models to fully obtain the weights through the exposed model fingerprint. We found that the second input to the last layer (i.e., the layer input) is subsequently small enough to fit into local memory. Thus, the entire layer input can be stolen, and the attacker can reproduce the final layer computation to uncover the final result of the DNN.

We note that this is a fairly straightforward attack, and with further creativity and ingenuity, a threat actor may be able to construct further complex and sophisticated malicious scenarios that may compromise ML applications in more severe ways. Below we provide a detailed description of the PoC, and the configuration and testing carried out on various GPU platforms to uncover their susceptibility to LeftoverLocals.

\paragraph{Our configuration}
%wizardLM-7B.ggmlv3.q5_0.bin
We outline our configuration in Tab.~\ref{tab:configuration}. Our attack builds on the llama.cpp LLM due to its simplicity and cross-vendor GPU acceleration support. In our example we use a large discrete GPU that we found to be susceptible to LeftoverLocals: the AMD Radeon RX 7900 XT. We configure llama.cpp to use OpenCL for GPU acceleration, which uses the CLBLAST linear algebra library\footnote{\url{https://github.com/CNugteren/CLBlast}}. We use the \texttt{wizardLM-7B.ggmlv3.q5\_0.bin} model, which is open source and can be obtained from Hugging Face\footnote{\url{https://huggingface.co/TheBloke/wizardLM-7B-GGML/tree/main}}. This model was selected due to its reasonable size, which enabled rapid prototyping and analysis; however, this attack is transferable to many different models. In our threat model, we assume that the victim is using the LLM in an interactive chat session.

\begin{table}[]
\centering
\small
       \caption{The details of our PoC: code base and system}
       \begin{tabular}{l l}
    \toprule
       LLM code base & llama.cpp \\
       Model & \texttt{wizardLM-7B.ggmlv3.q5\_0.bin} \\
       Layers & 33 \\
       GPU backend & OpenCL \\
       GPU BLAS Library & CLBLAST \\
       GPU & AMD Radeon RX 7900 XT \\
       Total local memory & 5MB \\
       \bottomrule
       \end{tabular}
       \label{tab:configuration}
\end{table}

\paragraph{Modification} The attack requires an optimized GPU implementation of matrix-vector multiplication. We found that the current matrix-vector multiplication in llama.cpp (which does not call into CLBLAST) is not implemented in an optimized idiomatic way. It stores partial dot product results in local memory and then combines them at the end. While there is a more complex approach using linear algebra to achieve our same results, for the simplicity of our PoC and demonstration, we replace the llama.cpp matrix-vector multiplication with our own that is more idiomatic (following best GPU programming programming practices).

\paragraph{Step 1—Fingerprinting the model} An attacker can fingerprint a model if it can listen to several inference queries from the victim. In our configuration, the GPU contains roughly 5MB of local memory. The model has roughly 33 layers, each of them consisting of a matrix multiplication operation. Matrix multiplication is often optimized on GPUs by using tiling: an approach that subdivides the matrices into small matrices, performs the multiplication, and then combines the results\footnote{\url{https://cnugteren.github.io/tutorial/pages/page4.html}}. In many optimized libraries, including CLBLAST, local memory is used to cache the smaller matrices. Thus, for every layer, the attacker can steal ~2.5MB of weights, and ~2.5MB of the inputs. While this is a significant amount of data, we note that it is not enough to reconstruct the entire computation. Many of these layers have weights and inputs that are 100s of MB large.

However, for a whole inference computation (33 layers), the attacker can steal around 80MB of the weights, which is sufficient to fingerprint the model (assuming the user is using an open-source model, such as one that can be found on Hugging Face). Given this, we assume that it is a straightforward task to fingerprint the model, and thus for the attacker to obtain the full model being used by the victim.

\paragraph{Step 2—Listening to the LLM output} The attacker can then turn their attention to the output layer of the DNN. In our configuration, we found that the output layer is a matrix-vector multiplication, rather than a matrix-matrix multiplication. The weights matrix is large (~128MB), but the input vector is quite small (~4KB). However, given that the attacker has fingerprinted the model in step 1, the attacker does not need to comprehensively steal the weights as they are available from the fingerprinted model.

Matrix-vector multiplication has a different GPU implementation than matrix-matrix multiplication. In the case where the input vector fits in local memory, the most performant implementation is often to cache the input vector in local memory, as it is used repeatedly (i.e., for repeated dot products). Because the input vector is stored entirely in local memory, the attacker can steal this entire vector. In determining whether the attacker has found local memory from the output layer, we discovered that the attacker could simply look for 4KB of floating point values with zeros on either side. In our testing, this unique fingerprint was associated with the output layer nearly every single time. For different models and different GPUs, this fingerprint will likely have to be recalibrated.

\paragraph{Putting it together} With an attacker in possession of both the weights and the input vector to the final layer of the DNN, they can perform the matrix-vector multiplication and obtain the resulting token returned by the inference. This allows the attacker to reproduce the output of the victim’s LLM chat session with high fidelity, as demonstrated in the introduction. In practice, we tuned the attacker to dump the local memory very efficiently (that is, by using only a small number of work-items and allocating only a small amount of memory). This allows the attacker to listen to long chat queries with only a small number of noticeable output artifacts. Some of the artifacts observed include:

\begin{itemize}
\item \textit{Duplicate tokens}: This occurs when the attacker steals the same output layer twice due to circumstances such as the attacker process being scheduled twice in a row, thus the LLM was not scheduled to compute its next token.
\item \textit{Missing tokens}: This occurs when the attacker kernel isn’t scheduled at the right time, i.e., immediately after the output layer computation kernel.
\item \textit{Incorrect tokens} This occurs due to
the attacker mis-identifying a stolen set of data to be the last layer. In this case, it will print a junk token.
\item \textit{Similar tokens} These tokens are “close” to the original output, even if they are not exact. That is, the attacker may be unable to steal the exact token embedding at the target layer. This results in a corrupted token embedding which, when decoded, is semantically similar (in the word2vec sense) to the original token. As an example, in Fig. \ref{fig:LLM-leak}, the attacker extracts the incorrect word “Facebook”, which is semantically similar to other Named Entities tokens (like “Google”, and “Amazon”) in the generated text.
\end{itemize}

Despite these discrepant artifacts, the stolen text is more than sufficient to uncover the LLM response. Additionally, the attacker can be further tuned by, for example, having multiple threads launch the listener kernel or by having a more precise fingerprint of the last layer.

\section{Testing GPU platforms for LeftoverLocals \label{sec:testing}}

Given the diversity of the GPU devices, we wrote several applications to test for LeftoverLocals, provided in a variety of GPU programming frameworks:

\begin{itemize}
    \item \textbf{Vulkan Command Line}: A command line application using Vulkan. The kernel is written in OpenCL and compiled to SPIR-V using \texttt{clspv}\footnote{\url{https://github.com/google/clspv}}. It uses a simple Vulkan wrapper called EasyVK\footnote{\url{https://github.com/ucsc-chpl/easyvk}}.
\item \textbf{OpenCL Command Line}: A command line application that uses the OpenCL framework.
\item \textbf{Apple App}: An Apple app that can be deployed on iOS or Mac OS. It targets the GPU using Apple’s Metal framework.
\item \textbf{Android App}: An Android app that uses Vulkan to target mobile GPUs. The code uses Vulkan’s C API (through EasyVK again) using JNI. The kernels are the same as in the Vulkan command line app: they are written in OpenCL and compiled to SPIR-V using \texttt{clspv}.
\end{itemize}

Using the above programs, we tested 11 devices spanning seven GPU vendors (and multiple GPU frameworks in some cases). We observed LeftoverLocals on devices from three of the vendors (Apple, Qualcomm, and AMD). The amount of memory leaked depends on the size of the GPU. Larger GPUs contain more physical memory, and thus, leak more data. For the larger GPUs (e.g., an AMD Radeon RX 7900 XT), we found that we can leak over ~5MB per kernel. Table~\ref{tab:leftover-positive} shows the system information in the cases where we were able able to observe LeftoverLocals (QC refers to Qualcomm):

\newcolumntype{b}{X}
\newcolumntype{s}{>{\hsize=.4\hsize}X}
\newcolumntype{f}{>{\hsize=.8\hsize}X}

\begin{table*}
\centering
%\footnotesize
\caption{GPUs and systems that we observed to be impacted by LeftoverLocals. QC refers to Qualcomm. \label{tab:leftover-positive}}
\begin{tabularx}{\textwidth}{fsb}
\toprule
\textbf{Device (GPU)} & \textbf{Framework} & \textbf{OS/Driver/Build system} \\
\midrule
  Apple iPhone 12 Pro (A14) & Metal & iOS 16.6, Xcode 14.3.1 (14E300c) \\ 
        Apple iPad Air 3rd G (A12) & Metal & iOS 16.5.1, Xcode 14.3.1 (14E300c) \\ 
        Apple MacBook Air (M2) & Metal & MacOS 13.4.1, Xcode 14.3.1 (14E300c) \\ 
        AMD Radeon RX 7900 XT & Vulkan & Arch Linux, Mesa 23.1.4 \\ 
        AMD Radeon RX 7900 XT & OpenCL & Arch Linux, OpenCL 2.1 AMD-APP.dbg (3570.0) \\ 
        AMD Ryzen 7 5700G (integrated GPU) & Vulkan & Arch Linux, Mesa 23.1.4 \\ 
        AMD RX 6700 XT & Vulkan & Windows 11 Pro 22H2, AMD Vulkan 2.0.270 \\ 
        HTC 1+ 11 (QC Snapdragon 8 g2) & Vulkan & Android 13, Android Studio (2022.3.1) \\ 
        HTC 1+ 5T (QC Snapdragon 835) & Vulkan & Android 13, Android Studio (2022.3.1) \\ 
\bottomrule
\end{tabularx}
\end{table*}

\begin{table*}
\centering
%\footnotesize
\caption{GPUs and systems where the listener returned non--zero values, but we could not confirm that they came from a different process. Arm representatives reviewed the results and confirmed that these values did not come from different processes on their devices.\label{tab:leftover-maybe}}
\begin{tabularx}{\textwidth}{fsb}
\toprule
\textbf{Device (GPU)} & \textbf{Framework} & \textbf{OS/Driver/Build system} \\
\midrule
        Galaxy Tab A (Arm Mali G78) & Vulkan & Android 13, Android Studio (2022.3.1) \\ 
        Google Pixel 6 (Arm Mali G71) & Vulkan & Android 11, Android Studio (2022.3.1) \\ 
        Google Pixel 7 (Arm Mali G710) & Vulkan & Android 13, Android Studio (2022.3.1) \\ 
\bottomrule
\end{tabularx}
\end{table*}

\begin{table*}
\centering
%\footnotesize
\caption{GPUs and systems that we did NOT observe to be impacted by LeftoverLocals. However, later representatives from Google notified us that they observed other devices from Imagination (IM) were impacted, and have subsequently been patched, as documented in the paper introduction.  \label{tab:leftover-negative}}
\begin{tabularx}{\textwidth}{fsb}
\toprule
\textbf{Device (GPU)} & \textbf{Framework} & \textbf{OS/Driver/Build system} \\
\midrule
        Nvidia GeForce RTX 4070 & Vulkan & Arch Linux, Mesa 23.1.4 \\ 
        Nvidia GeForce RTX 4070 & OpenCL & Arch Linux, OpenCL 3.0 CUDA 12.2.128 \\ 
        Intel NUC (NUC10I5FNK) & Vulkan & Ubuntu 22.04, Mesa 20.3.2 \\ 
        Intel NUC (NUC10I5FNK) & OpenCL & Ubuntu 22.04, OpenCL 3.0 NEO (22.31.23852) \\ 
        Motorola M.G (IM PowerVR GE8320) & Vulkan & Android 11, Android Studio (2022.3.1) \\ 
\bottomrule
\end{tabularx}
\end{table*}

For some devices, specifically those from Arm, we were not able to observe the canary value from the writer in the listener, but we did observe non-zero data. Representatives from Arm reviewed our observations and concluded that although these values are not zero, they are not from a memory leak. We outline these devices in Tab.~\ref{tab:leftover-maybe}.

Additionally, we tested some GPUs from NVIDIA, Intel, and Imagination, outlined in Tab. \ref{tab:leftover-negative} (IM refers to Immagination). For these devices, we observed only zeros in local memory, and thus did not observe LeftoverLocals. It is unclear if all their devices are not impacted. For example, although we did not observe the issue on our Imagination device, Google notified us that they were able to observe it on other Imagination devices.

We prepared a YouTube video that demonstrates the different interfaces and examples of LocalLeftovers—namely the LLM PoC attack, covert communication channels, and searching for canary values—on a few different platforms using a few different applications\footnote{\url{https://www.youtube.com/watch?v=g2A7GvbnItg}}. All code is provided in our github repository.

\paragraph{Vulnerable environments} An attack program must be co-resident on the same machine and must be “listening” at the same time that the victim is running a sensitive application on the GPU. This could occur in many scenarios: for example, if the attack program is co-resident with the victim on a shared cloud computer with a GPU. On a mobile device, the attack could be implemented in an app or a library. Listening can be implemented efficiently, and thus can be done repeatedly and constantly with almost no obvious performance degradation.

Next, we briefly discuss other environments where GPUs are either deployed or where an attacker might have access to sensitive information. Although it appears that some current systems (e.g., WebGPU) are not currently impacted, the ever-growing prevalence of ML and the diversity of modern GPUs mean that the next iteration of these systems (or other near-future systems) may be severely compromised by these types of vulnerabilities.

\begin{itemize}
\item \textbf{Cloud providers}: Cloud providers (e.g., AWS and Azure) are unlikely to provide shared GPU instances, especially if users have dedicated access to the GPU machine. In other cases, GPUs could be shared using very conservative GPU VM technology (such as NVIDIA’s vGPU\footnote{\url{https://docs.nvidia.com/grid/13.0/grid-vgpu-user-guide/index.html}} or AMD's MxGPU\footnote{\url{https://www.amd.com/system/files/documents/gpu-consistency-security-whitepaper.pdf}}), which physically partitions the GPU and therefore prevents users from sharing GPU resources (including local memory). Given this, many current cloud GPU systems may not currently be vulnerable to LeftoverLocals; however, we do not have conclusive evidence to determine this given the general lack of visibility into the specification and implementation of these systems. We note that we have observed LeftoverLocals on multi-user Linux servers, as well as on desktop (Windows and Mac) systems through traditional multi-processing. This includes Docker containers on these systems.
\item \textbf{Mobile applications}: In our experiments and explorations in the mobile domain, we were able to run concurrent GPU processes (from different apps on iOS or Android) only in very specific instances. That is, we were not able to run a GPU process (e.g., from a malicious listener app) in the background while other apps (e.g., the victim) were run in the foreground. As with our analysis of cloud providers, we were unable to find clear documentation that explicitly detailed these constraints, and so we cannot definitively claim whether they are vulnerable. However, as seen our youtube video, LeftoverLocals can be exploited either when a malicious listener app is run side-by-side with a victim app, or if the malicious listener app is quickly swapped from the background into the foreground from a victim app.
\item \textbf{Remote attacks}: We preliminarily investigated the possibility of attacks originating from websites (e.g., those hosted by a remote attacker). To our knowledge, web applications do not have the low-level features required to listen to local memory using GPU graphics frameworks, such as WebGL. We note that the new WebGPU framework does provide low-level capabilities that allow a webpage to access local memory. Conservatively, WebGPU initializes and performs dynamic array bounds checking on local memory (and global memory), which mitigates this vulnerability. However, these checks cause significant overhead, as documented in online discussions\footnote{\url{https://github.com/gpuweb/gpuweb/issues/1202}}. To test this further, our code repo contains a simple listener in WebGPU. As expected, we have only observed zeros in local memory, even on devices that are vulnerable to LeftoverLocals through other frameworks. However, GPU compilers are known to be fragile~\cite{clsmith}, and it is not difficult to imagine finding a compiler bug that could somehow bypass these checks (especially using fuzzing techniques). Our position is that LocalLeftovers should be addressed at a lower level (e.g., the driver).
\end{itemize}

\paragraph{How GPU vendors can resolve this vulnerability} To defend against LocalLeftovers, GPUs should clear their local memory between kernel calls. While this could cause some performance overhead, our experiments show that many GPU vendors (e.g., NVIDIA, Intel) currently appear to provide this functionality. It even appears that some of this functionality is provided for impacted GPUs. For example, Mesa drivers for AMD GPUs clears local memory after a compute kernel launch\footnote{\url{https://github.com/Mesa3D/mesa/blob/957009978ef6d7121fc0d710d03bc20097d4d46b/src/amd/vulkan/radv_shader.c\#L709}}. However, this approach has a fundamental flaw that makes it vulnerable to LeftoverLocals: this memory wipe is done with a separate kernel, thus, the GPU kernel queue may contain a malicious listener between the computation kernel and the local memory wipe, allowing the listener to steal memory. Instead, the computation kernel and the local memory wipe need to occur atomically, i.e., without allowing any other kernel to be interleaved between them. Otherwise, a user may attempt to preemptively defend themselves against LeftoverLocals as described in the next section.

\paragraph{Mitigations} Given the lack of comprehensive patches across impacted GPU vendors, LeftoverLocals can be defended by modifying the source code of all GPU kernels that use local memory. As we’ve previously noted, before the kernel ends, the GPU threads can store 0 to any local memory locations that were used in the kernel. Given that GPU tasks are typically interleaved at the kernel boundary, this will prevent another user from being able to read leftover values. We note that this mitigation may be difficult for many users, especially because GPU code is often buried deep in complex software stacks (e.g., for ML). Furthermore, the GPU code may be part of a highly optimized library (e.g., ML linear algebra routines). In these cases, it is very difficult to identify how local memory is used, and even more difficult to modify the kernel to zero it out. It may be possible to augment a compiler to add this functionality, similar to how WebGPU handles GPU memory accesses. These mitigations do have a performance overhead that should be taken into account. Another blunt mitigation involves simply avoiding multi-tenant GPU environments.

\section{Impact on LLMs and GPU platforms}

\paragraph{LLM security}

Our PoC attack examines only one application: an interactive open-source LLM session. However, with a little creativity, attackers could likely target many GPU applications, including those used within privacy-sensitive domains. Our motivation stems from the recent increased use and support of open-source models, often accompanied by claims that their “openness” inherently entails safety and security through transparency. A recent article in Nature\footnote{\url{https://www.nature.com/articles/d41586-023-03803-y}} even alleges that only open-source generative AI models can “safely” revolutionize health care, a safety-critical domain. Yet, even if open-source models provide the opportunity to be rigorously audited and assessed (which they have yet to be\footnote{\url{https://blog.trailofbits.com/2023/11/15/assessing-the-security-posture-of-a-widely-used-vision-model-yolov7/}}), their deployment still hinges on a closed-source stack (i.e., GPUs). And as demonstrated by LeftoverLocals, open-source LLMs are particularly susceptible to our vulnerability given our ability to fingerprint these models to obtain remaining weights as needed. Indeed, we have already observed announcements regarding the deployment of open-source models in collaboration with impacted GPU vendors, including Hugging Face’s collaboration with AMD\footnote{\url{https://twitter.com/AMD/status/1744831880241750112}}, Lamini’s deployment on AMD GPUs\footnote{\url{https://www.lamini.ai/blog/lamini-amd-paving-the-road-to-gpu-rich-enterprise-llms}}, and the Qualcomm and Meta partnership for edge devices \footnote{\url{https://www.qualcomm.com/news/releases/2023/07/qualcomm-works-with-meta-to-enable-on-device-ai-applications-usi}}.

Generally, the introduction of ML poses new attack surfaces that traditional threat models do not account for, and that can lead to implicit and explicit access to data, model parameters, or resulting outputs, increasing the overall attack surface of the system. It is crucial to identify and taxonomize novel classes of failure modes that directly impact ML models, in addition to novel threats that can compromise the ML Ops pipeline, as we have demonstrated with LeftoverLocals. We discuss GPU-specific threat implications in the following section.

\paragraph{GPU providers, applications, and vendors}

While many platforms are not currently impacted, we emphasize that the GPU compute landscape is evolving rapidly. As some examples: there are a growing number of GPU cloud providers have various policies and available configurations\footnote{\url{https://cloud-gpus.com/}}; and GPU programming frameworks, such as Vulkan and Metal, are well-supported on mainstream platforms, and can be used in apps without requiring extra privileges. While these developments are exciting, they increase the threat potential of GPU vulnerabilities, as LeftoverLocals illustrates. As far as we are aware, there is no unified security specification for how GPUs are required to handle sensitive data, and no portable test suite to check if systems are vulnerable to simple memory leaks, like LeftoverLocals. Thus, GPU compute environments should be rigorously scrutinized when used for processing any type of sensitive data.

As mentioned throughout this paper, while we focus on LLM applications, GPU local memory is one of the first tools that a GPU developer uses when optimizing an application. Although these attacks require analyzing the victim’s GPU kernel code to identify local memory usage, it is likely possible to find similar application attacks in other GPU compute domains, such as image processing and scientific computing. It will be increasingly difficult for users to detect and defend against these attacks since it’s unlikely they will know if their application is vulnerable to LeftoverLocals; this would require knowing the details of the exact GPU kernel code, which are often hidden away in highly optimized linear algebra libraries (e.g., CLBLAST). Additionally, an overall lack of specification in up-and-coming GPU platforms makes it difficult to determine whether the compiler or runtime will use impacted memory regions without the user knowing. For example, Apple GPUs have a new caching mechanism, called dynamic caching\footnote{\url{https://www.digitaltrends.com/computing/apple-dynamic-caching-explained/}}, which does not have a clear specification regarding if local memory regions are being used for other purposes.

\section{Coordinated disclosure}

Since September 2023, we have been working CERT/CC on a large coordinated disclosure involving all major GPU vendors, including NVIDIA, Apple, AMD, Arm, Intel, Qualcomm, and Imagination. Trail of Bits provided vendors a total of 125 days to test their products and provide remediations. The coordination gradually grew to include software stakeholders, including Google, Microsoft, and others, which allowed us to understand how LocalLeftovers impacts privacy requirements and impact at different stages in the ML supply chain. Apple did not respond or engage with us regarding the disclosure until three days before the embargo ended.

A high-level timeline of the disclosure is provided below:

\begin{itemize}
\item September 8, 2023: Trail of Bits submitted report to the CERT/CC
\item September 11, 2023: CERT/CC acknowledged the submission of LeftoverLocals and began the process of vendor outreach and CVE assignment with a preliminary disclosure date of December 11, 2023
\item September 14, 2023: AMD acknowledged the CERT disclosure
\item September 15, 2023: Qualcomm acknowledged the CERT disclosure
\item September 22, 2023: The case report was shared with the Khronos group, which produces the OpenCL and Vulkan specifications and some corresponding infastructure (such as conformance testing)
\item September 29, 2023: NVIDIA acknowledged disclosure and confirmed they were not affected by the vulnerability
\item November 22, 2023: ToB extended release of embargo to January 16, 2024 to accommodate for vendor requests for further time
\item January 11, 2024: We received a notice that Qualcomm provided a patch to their firmware that addresses this issue only for some of their devices. Additionally, Google noted that ChromeOS Stable 120 and LTS 114 will be released on January 16 to include AMD and Qualcomm mitigations.
\item January 13, 2024: Apple confirmed that the A17 and M3 series processors contain fixes to the vulnerability.
\item January 14, 2024: Google notified us that they observed that that some Imagination GPUs are impacted.
\item January 16, 2024: Embargo lift and public disclosure of LeftoverLocals
\end{itemize}

\section{Moving forward}
Now that GPUs are being used in a wide range of applications, including privacy sensitive applications, we believe that the wider GPU systems community (vendors, researchers, developers) must work towards hardening the GPU system stack and corresponding specifications. This should be accomplished through robust, holistic specifications that describe both GPU programs’ behavior and how GPU devices integrate with the rest of the system stack (e.g., the OS or hypervisor). Furthermore, these specifications should be rigorously tested to account for the diversity of GPU systems and safety requirements of diverse application domains. Looking forward, a wide variety of new AI chips are being developed\footnote{\url{https://www.theinformation.com/articles/the-twelve-startups-battling-for-a-slice-of-nvidias-pie?utm_source=ti_app}} and will require rigorous security analysis.

There are positive developments in this direction. For example, AMD’s ROCm stack\footnote{\url{https://www.amd.com/en/products/software/rocm.html}} is open, and thus available for independent rigorous evaluation, and the Khronos Group has safety critical specification groups\footnote{\url{https://www.khronos.org/syclsc}}. Additionally, cross-vendor programming frameworks, such as Vulkan, have been incredibly useful for writing portable test suites, as opposed to single-vendor programming frameworks.

While GPU security and privacy guarantees are scattered and scarce, the Vulkan specification outlines a reasonable definition of security for GPU platforms to adhere to—a definition that several platforms clearly violate, as LocalLeftovers shows:\footnote{\url{https://registry.khronos.org/vulkan/specs/1.3-extensions/html/vkspec.html\#fundamentals-validusage}}

\begin{quote}
\ldots implementations must ensure that [\ldots] an application does not affect the integrity of the operating system [\ldots]. In particular, any guarantees made by an operating system about whether memory from one process can be visible to another process or not must not be violated by a Vulkan implementation for any memory allocation. 

\end{quote}

With the dust settling, our position is the following: given the wide diversity of GPUs and their critical importance in enabling machine learning applications, these devices, and their ecosystems, are in need of (1) a detailed threat model that considers the various types of data processed on GPUs and how this data might be compromised; (2) an exploration of the GPU execution stack to determine where and how GPU security properties should be specified and implemented; and (3) significant testing and auditing to fortify GPU ecosystem, which is the computational foundation of machine learning.

\textit{For full transparency, we note that Tyler Sorensen has been an invited member of the Khronos group (sponsored by Google) since 2019, and participates in the memory model technical specification group.}

\section*{Acknowledgements} We thank Dan Guido, Trent Brunson, Max Ammann, Dominik Czarnota, Kelly Kaoudis, Jay Little, and Adelin Travers for their insightful comments and feedback on the vulnerability, PoC, and throughout the disclosure process. We also thank the Khronos Group for discussing technical specification details with us, and providing an avenue for us to engage with many vendors. We thank CERT/CC, specifically Vijay Sarvepalli and Ben Koo, for organizing the coordinated disclosure, especially considering the potential breadth of the vulnerability. Finally, thank you to everyone who engaged with us on this issue. This was a large project and we had discussions with many people who provided valuable insights and perspectives.

\bibliographystyle{ACM-Reference-Format}
\bibliography{references}

\end{document}